\documentclass[prl,aps,amssymb,nofootinbib,twocolumn]{revtex4}
\usepackage{amsmath}
\usepackage{amssymb}
\usepackage{amsthm}
\usepackage{amsfonts}
\usepackage{algorithmic}
\usepackage{enumerate}
\usepackage{latexsym}
\usepackage[dvips]{graphicx}

\newcommand{\beq}{\begin{equation}}
\newcommand{\eneq}{\end{equation}}
\newcommand{\beqs}{\begin{equation*}}
\newcommand{\eneqs}{\end{equation*}}

\input{epsf}

\begin{document}

\tolerance 10000

\title{On The Existence of Roton Excitations in Bose Einstein
Condensates: Signature of Proximity to a Mott Insulating Phase}

\author { Zaira Nazario$^\dagger$ and
David I. Santiago$^{\dagger, \star}$ }

\affiliation{ $\dagger$ Department of Physics, Stanford
University,Stanford, California 94305 \\ 
$\star$ Gravity Probe B Relativity Mission, Stanford, California 94305}

\begin{abstract}
\begin{center}

\parbox{14cm}{Within the last decade, artificially engineered Bose
Einstein Condensation has been achieved in atomic systems. Bose Einstein
Condensates are
superfluids just like bosonic Helium is and all interacting bosonic fluids
are expected to be at low enough temperatures. One difference between the two
systems is that superfluid Helium exhibits roton excitations while
Bose Einstein Condensates have never been observed to have such
excitations. The reason for the roton minimum in Helium is its
proximity to a solid phase. The roton minimum is a consequence of
enhanced density fluctuations at the reciprocal lattice vector of the
stillborn solid. Bose Einstein Condensates in atomic traps are not
near a solid phase and therefore do not exhibit roton minimum. We
conclude that if Bose Einstein Condensates in an optical lattice are
tuned near a transition to a Mott insulating phase, a roton minimum
will develop at a reciprocal lattice vector of the
lattice. Equivalently, a peak in the structure factor will appear at
such a wavevector. The smallness of the roton gap or the largeness of the 
structure factor peak are experimental signatures of the proximity to the Mott 
transition.}

\end{center}
\end{abstract}

\date{\today}

\maketitle

In the present work we focus attention in the possible existence or
nonexistence of roton excitations in artificially engineered Bose
Einstein Condensates (BECs)\cite{beca,becb,becc}. BECs are superfluid
as they posses a finite sound speed\cite{bec2,landau,bog} and reduced
long wavelength scattering\cite{bec3,bec4}. The other important
bosonic superfluid is He$_4$\cite{landau,pines}. Since BECs and He$_4$
are the same universal phase of matter, they share a large
commonality. For example, they are both superfluids whose elementary
long wavelength excitation spectrum is phononic, their ground states  
spontaneously break gauge invariance\cite{phil}, that is their ground states 
exhibit macroscopic quantum coherence, and they both posses
vortex excitations\cite{becv}. This is an example of universality of
stable fixed points of matter. In these cases the superfluid ground
state is such a fixed point. The similarities will become ever more apparent 
as the low energy properties BECs are studied more carefully.

On the other hand, at shorter wavelengths Helium possesses a roton
minimum\cite{landau,bog,feyn1} in its excitation spectrum leading
to low energy excitations occurring at a specific wavevector, which are
gapped and are different than long wavelength sound. These excitations
can make important contributions to the dynamics and thermodynamics of
the system. In BECs no such excitation has been observed and the lack
of a peak in the structure factor\cite{bec3} leads to the conclusion
that they do not posses roton excitations. We review some of the
properties of rotons in Helium and determine under what conditions
will they occur in BECs.

All quantum many particle systems are described by a Hamiltonian\cite{landau}
\beq\label{haml}
\mathcal{H} = \int \left( m \frac{\vec v \cdot \rho \vec v}{2} \right)
d^3r + U[\rho]
\eneq
where $m \vec v \cdot \rho \vec v / 2$ is the kinetic energy operator,
$U[\rho]$ is the potential energy operator, which in general can be a
functional of the density operator. The density operator in first
quantized notation is
\beq
\rho(\vec r) = \sum_i \delta(\vec r - \vec r_i)
\eneq
with $\vec r_i$ being the position of the ith particle. The velocity 
operator is in first quantized notation
\beq
\vec v(\vec r) = \sum_i \left[ \frac{\vec p}{2m}\delta(\vec r - \vec r_i) +
 \delta(\vec r - \vec r_i)\frac{\vec p}{2m}\right]
\eneq
with commutation relations 
\beq \label{com}
\left[ \rho(\vec r\;')\;, \vec v(\vec r) \right] = i \frac{\hbar}{m} \nabla 
\delta(\vec r\;' - \vec r)
\eneq
The ground state of the Hamiltonian (\ref{haml}) could be a
quantum fluid or solid depending on the interaction.  We will
suppose it to be a fluid as we have BECs and Helium in mind.

We study the excitation spectrum of the bosonic fluid on quite general
grounds following Landau\cite{landau} and the Russian school\cite{abri} quite 
closely. We suppose the density to have a well defined average which
we take to be uniform for simplicity. This will not change the nature
of the considerations, although in real life the density can be
modulated by the lattice as for BECs in optical lattices. 
The average velocity is zero as we consider a system at rest. We Fourier expand
the density and velocity operators about their averages:
\begin{align}
\rho(\vec r) &= \rho_0 + \frac{1}{N} \sum_{\vec k} \rho_{\vec k}
e^{i\vec k \cdot \vec r} \\
\vec v(\vec r) &= \frac{1}{N} \sum_{\vec k} v_{\vec k} e^{i\vec k \cdot
\vec r}
\end{align}
where we are using the lattice normalization, the number of
sites $N$ instead of the volume $V$. Quantum mechanically, the
velocity is proportional to the gradient of the local phase variable
of the particle, $\vec v(\vec r)=\hbar/m\nabla\theta$, or in Fourier
components
\beq
\vec v_{\vec k}=\frac{i\hbar\vec k}{m} \theta_{\vec k}
\eneq
\noindent The density velocity commutation relation (\ref{com}) imply
the well known density phase commutation relation
\beq
[\rho_{\vec k}\; , \; \; \theta_{-\vec k'}] = i \delta_{\vec k \vec k'}
\eneq
\noindent i.e. $\hbar\theta_{-\vec k}$ is the momentum conjugate to
the density fluctuation $\rho_{\vec k}$. The Hamiltonian (\ref{haml})
thus becomes 
\beq \label{ham2} 
\mathcal H = U(\rho_0) + \frac{1}{N} \sum_{\vec k} \left\{
\frac{\rho_0\hbar^2 k^2}{2m}|\theta_{-\vec k}|^2 + \frac{1}{2}
\left(U_{\vec k} + \frac{\hbar^2 k^2}{4m\rho_0}\right) |\rho_{\vec
k}|^2 \right\}
\eneq
\noindent where $U_{\vec k}$ is the Fourier transform of the
interaction. The interaction is completely general. The only
requirements are that $U_{\vec k=0}$ is a constant, so that we can
have a superfluid, and that there is some short range repulsion
analogous to a Hubbard term that can be turned up in order to
stabilize a Mott insulating phase. $(\hbar^2
k^2)/(4m\rho_0)|\rho_{\vec k}|^2$ is the elastic energy associated
with changing the density. This last term comes from
acting the Hamiltonian on the wavefunction and making sure that one
keeps careful track of both density and phase degrees of
freedom\cite{dhl2} in the full quantum mechanical kinetic energy term. We
have performed studies very near to these in the context of fermionic 
systems\cite{us1} which we follow very closely in the present note. Charged 
boson systems were studied previously\cite{dhl2} with the Quantum Hall effect 
in mind.

The quantum liquid is thus a collection of harmonic oscillators
in momentum space for the density fluctuations. The mass and spring constant 
of the oscillators are 
\beq
M_{\vec k} \equiv \frac{m}{\rho_0 k^2} \; , \; \; K_{\vec k} \equiv U_{\vec k} 
+ \frac{\hbar^2 k^2}{4m\rho_0} \; .
\eneq 
\noindent The density energy excitation spectrum of the quantum liquid
is
\beq
E_{\vec k} = \hbar \omega_{\vec k} (n + 1/2) 
\eneq
\beq 
\omega_{\vec k}^2 = \frac{K_{\vec k}}{M_{\vec k}} = \frac{\rho_0
k^2}{m} \left(U_{\vec k} + \frac{\hbar^2 k^2}{4m\rho_0} \right)
\eneq
\noindent The ground state energy is given by $U(\rho_0) + \sum_{\vec k} \hbar
\omega_{\vec k} / 2$. We note that the Virial theorem implies that
\beq \nonumber
\frac{1}{2} \hbar \omega_{\vec k} N = \left( U_{\vec k} +
\frac{\hbar^2 k^2}{4m\rho_0} \right) \langle |\rho_{\vec k}|^2 \rangle
\eneq
\beq \label{ex}
\hbar \omega_{\vec k} = \frac{\hbar^2 k^2}{2 m S(\vec k)}
\eneq
\noindent where the structure factor is defined by 
\beq \label{s}
S(\vec k) =
\frac{\langle |\rho_{\vec k}|^2 \rangle}{N \rho_0} 
\eneq
\noindent which is, of course, the Fourier transform of the
density-density correlation function. For the quantum liquid we thus have
\beq 
S(\vec k) =\frac{\hbar k}{2m}\frac{1}{\sqrt{\frac{\rho_0}{m} \left(U_{\vec k}+
\frac{\hbar^2 k^2}{4m\rho_0}\right)}}
\eneq
\noindent The ground state energy will have the density oscillators unexcited.
Hence, if the ground state wavefunction of the bosonic system 
{\it without density correlations}  is $|\psi_{GS}\rangle$, the ground state 
wavefunction {\it with density correlations} is
\beq \label{gs}
|\psi\rangle = \exp\left \{ -\frac{1}{2}\sum_{\vec k} 
\sqrt{\frac{m}{\rho_0 \hbar^2 k^2 }
\left(U_{\vec k} + \frac{\hbar^2 k^2}{4 m \rho_0}\right)} |\rho_{\vec
k}|^2 \right\}|\psi_{GS} \rangle \; .
\eneq
\noindent The factor in front appears because it is the ground state 
wavefunction of harmonic oscillators and the minimum energy condition is 
obviously to have the oscillators shaking as little as possible. We see that 
the factor suppresses density fluctuations. The wavefunction is essentially 
exact to quadratic order in the density fluctuations. 

Since the Hamiltonian and ground state wavefunction do not contain
any periodic lattice effects, they aptly describe systems with no
external lattice such as liquid Helium, to which we turn our
discussion now. The ground state wavefunction (\ref{gs}) does not
contain any insulating crystalline order as $\langle \rho_{\vec k}
\rangle = 0$ for all $\vec k$. When the system spontaneously
solidifies, it will exhibit Bragg peaks at the reciprocal lattice vectors, 
$\vec Q$, corresponding to the insulating crystal. These peaks mean that
$\langle \rho_{\vec Q} \rangle \neq 0$. We inquire into the behavior
when the boson fluid approaches a solidification transition.

Proximity to a solid phase with reciprocal lattice vector $\vec Q$ for
the incipient order softens the spring constant of the density
oscillators of the liquid, $\vec K_{\vec Q} \rightarrow 0$, which
makes $\omega_{\vec Q} \rightarrow 0$. This last condition means the
mode becomes degenerate with the ground state and there will be
enhanced density fluctuations, $\langle |\rho_{\vec Q}^2| \rangle
\rightarrow \infty$ at the wavevector of the stillborn solid, i.e. the
system is approaching a phase transition. The diverging fluctuations
follow because the position fluctuations of a harmonic oscillator
diverge as the spring constant vanishes due to the Heisenberg
uncertainty principle. The softening of the density mode implies that
the interaction $U_{\vec k}$ has a minimum at $\vec Q$ which in turn
produces a peak in the structure factor, $S(\vec Q)$, that diverges as
one approaches the point when the system solidifies. This peak in the
structure factor implies will drag the spectrum down according to the
equation (\ref{ex}) creating a roton minimum\cite{feyn1}. 
\begin{figure}[ht]
\centering 
\rotatebox{270}{\resizebox{4.72cm}{!}{%
\includegraphics*{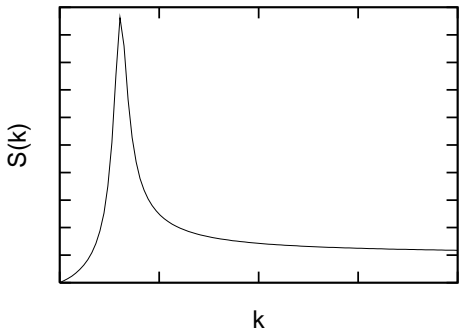}}}
\caption{Structure factor vs. wavevector for a Bose fluid near a
solidification transition. The density fluctuation peak at the
wavevector of the incipient order is clearly visible.}
\end{figure}
\begin{figure}[ht]
\centering 
\rotatebox{270}{\resizebox{4.72cm}{!}{%
\includegraphics*{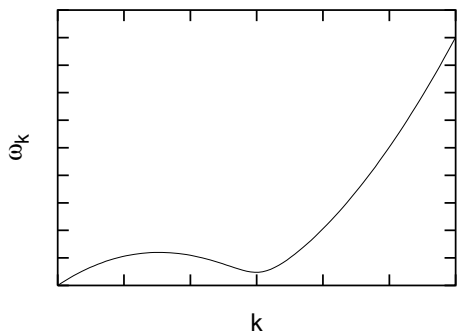}}}
\caption{Frequency dispersion for a Bose fluid near a solidification
transition. The roton minimum at the wavevector of the incipient order
is clearly visible.}
\end{figure}

We thus have as a simple phenomenological model that captures the
physics
\beq
\nonumber U_{\vec k} = \frac{U}{k^2 + k_0^2} - \frac{U}{Q^2 + k_0^2} +
\alpha \frac{\hbar^2}{4 m \rho_0} (\vec k - \vec Q)^2 - \alpha
\frac{\hbar^2 k^2}{4 m \rho_0} + \Delta
\eneq
\noindent where $k_0$ is some microscopic scale that characterizes the
interaction, nonzero $\alpha$ makes a roton minimum, and
$\Delta-\alpha \frac{\hbar^2 Q^2}{4 m \rho_0}$ is the roton gap, which
will collapse at the crystallization transition as the smallness of
the gap of such a roton is a measure of proximity to
crystallization. The first term in the potential represents the short
range repulsion, chosen for simplicity to have a Yukawa form. The
parameter $\Delta$ can be thought of as the long range repulsion. From
the form of the interaction we get the structure factor and the
density excitation spectrum for Helium, which we plot in Figures 1 and
2. We emphasize that the reason for the roton minimum in the
excitation spectrum is the proximity of the liquid Helium ground state
to a solid phase. Helium is an almost solid barely melted by quantum
fluctuations, i.e. it is a very strongly correlated liquid.

We now turn our attention to BECs in optical lattices. These
superfluids do not exhibit roton minimum in their excitation
spectrum. The reason for this is that unless the wells of the optical
lattice are made deep, the system is far from a solid or Mott
insulating phase. In fact, it is an extremely good approximation to
take the interaction potential to be a constant $U_{\vec k} = I$. For
such a case, we plot in Figures 3 and 4 the structure factor and
density excitation spectrum. We see no peak in the structure factor or
roton minimum in the excitation spectrum as observed
experimentally\cite{bec3}.
\begin{figure}[ht]
\centering 
\rotatebox{270}{\resizebox{4.72cm}{!}{%
\includegraphics*{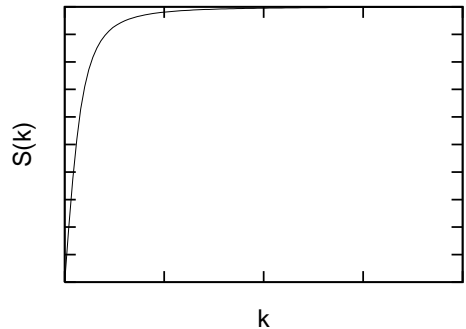}}}
\caption{Structure factor vs. wavevector for a Bose fluid far from a
solidification transition. There is no peak.}
\end{figure}
\begin{figure}[ht]
\centering 
\rotatebox{270}{\resizebox{4.72cm}{!}{%
\includegraphics*{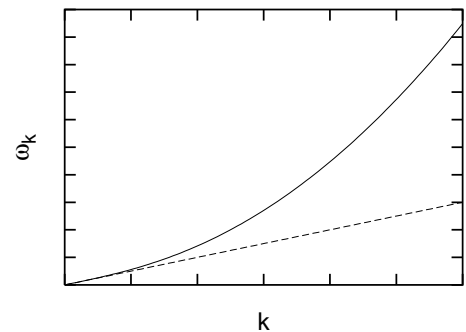}}}
\caption{Frequency dispersion for a Bose fluid far from a
solidification transition. The upper graph is the superfluid
dispersion and the lower one is a phononic linear dispersion in order
to make aparent the linearity of the superfluid dispersion at long
wavelengths. There is no roton minimum.}
\end{figure}

So far we have neglected optical lattice effects because even though
they make quantitative changes to measurable parameters of the BEC
superfluid, they do not change the universal superfluid physics. On
the other hand, as one approaches a commensurate Mott insulating
phase, the lattice will become ever more important. The presence of
the optical lattice implies that there is an external potential $-
V_{\vec k}$ that acts on the bosons thus adding to the Hamiltonian (\ref{ham2})
the term
\beq
\label{hamext} \mathcal H_{\text{ext}} = - \sum_{\vec k} V_{\vec k}
\rho_{\vec k} \; .  
\eneq 
The effect of the extra term in the ground state wavefunction is to
move the equilibrium position of the density oscillators from $\langle
\rho_{\vec k} \rangle = 0$ to
\beq
\label{rho} \langle \rho_{\vec k} \rangle = \frac{V_{\vec k}}{U_{\vec
k} + \frac{\hbar^2 k^2}{4m\rho_0}} \; .  
\eneq

The fact that there is $\langle \rho_{\vec k} \rangle \neq 0$ {\it does not} 
mean the boson fluid {\it in an optical lattice} has solidified into 
an insulating phase. It just means that the boson fluid is being modulated by 
the periodic lattice. It is true that in the insulating phase this
expected value will be nonzero too. If there was not an external
lattice, $\rho_{\vec k} \neq 0$ for some $\vec k$'s would imply an
insulating phase as the only way this can happen is for the system to
spontaneously break translational invariance. So the question of how
to determine if one is in the Mott phase is delicate for lattice boson
systems. 
\begin{figure}[ht]
\centering 
\rotatebox{270}{\resizebox{4.72cm}{!}{%
\includegraphics*{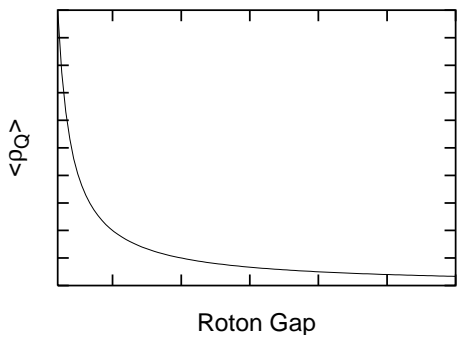}}}
\caption{Density Fourier component at the wavevector of the incipient
order vs. roton gap for a Bose fluid near a solidification
transition. The divergence as the roton gap goes to zero is a
signature of the imminent transition.}
\end{figure}

A way to know that the bosons in an optical lattice have solidified into a 
commensurate Mott phase is to look at the density excitation spectrum or the 
structure factor. When it has solidified, the excitation spectrum will have a
gap at long wavelengths and the structure factor will go like $k^2$ at
long wavelengths. This is not a violation of Goldstone's
theorem\cite{gold} as the lattice was not generated spontaneously by
the boson system but is provided by the external lattice potential,
i.e. the symmetry is explicitely broken. Our interest is what are the
signatures of the incipient Mott insulating phase. As mentioned above,
one can know the transition is approaching by a softening of the
excitation spectrum at appropriate wavevectors. In this case, the
system will behave exactly as Helium, exhibiting roton minimum  and its
structure factor and energy excitation spectrum will be as shown in
Figures 1 and 2. Moreover, from the formula for the response of the
lattice potential (\ref{rho}), we see that the density Fourier
component at the roton wavevector will grow as the roton gap
collapses. We plot the growth of such a Fourier component vs. size of
the roton gap in Figure 5.

Summarizing, artificially engineered BECs and liquid He$_4$ are both
quantum bosonic superfluids. They share a large similarity of
intrinsically quantum behavior for their ground states and low energy
states.  One important difference is that Helium possesses roton
excitations while BECs appear not to have them. The reason for this is
that BECs in the atomic traps are far from being a solid. Helium
posses a roton because it is a stillborn solid where quantum
fluctuations barely prevent solidification.

On the other hand, BECs in optical lattices\cite{ib,fisher,us} can be tuned 
near a solidification or Mott transition by making the wells deeper. 
We thus predict that near a transition from a superfluid BEC to a commensurate 
Mott phase in an optical lattice there will be roton excitations
measurable by light scattering occurring at the reciprocal lattice
vector of the lattice. Their experimental signatures will be a large peak in 
the structure factor or a roton minimum in the excitation spectrum.  The 
smallness of the roton gap or height of the structure factor peak can be used 
as diagnostics of the proximity to the incipient Mott order. The roton
excitation adds to the list of intrinsically quantum mechanical
excitations that exists in BECs and will be of importance to the dynamics 
of the systems.

{\bf Acknowledgments:} We thank Dung-Hai Lee for very inspiring
discussions. Zaira Nazario is a Ford Foundation predoctoral fellow.
She was supported by the Ford Foundation and by the School of
Humanities and Science at Stanford University. David I. Santiago was
supported by NASA Grant NAS 8-39225 to Gravity Probe B.

\end{document}